\documentclass[aps,pre,preprint]{revtex4}

\usepackage{amsmath}
\usepackage{amssymb}
\usepackage{colordvi,xcolor}
\usepackage[utf8]{inputenc}
\usepackage{graphicx}
\usepackage{dcolumn}
\usepackage[mathscr]{euscript}
\usepackage{bm}

\begin{document}

\title{Dynamics and kinetic theory of hard spheres under strong confinement}
\author{J. Javier Brey}
\email{brey@us.es}
\affiliation{F\'{\i}sica Te\'{o}rica, Universidad de Sevilla,
Apartado de Correos 1065, E-41080, Sevilla, Spain}
\affiliation{Institute for Theoretical and Computational Physics. Facultad de Ciencias. Universidad de Granada, E-18071, Granada, Spain}

\author{M. I. Garc\'{\i}a de Soria}
\email{gsoria@us.es}
\affiliation{F\'{\i}sica Te\'{o}rica, Universidad de Sevilla,
Apartado de Correos 1065, E-41080, Sevilla, Spain}
%\affiliation{Institute for Theoretical and Computational Physics. Facultad de Ciencias. Universidad de Granada, E-18071, Granada, Spain}
\author{P. Maynar}
\email{maynar@us.es}
\affiliation{F\'{\i}sica Te\'{o}rica, Universidad de Sevilla,
Apartado de Correos 1065, E-41080, Sevilla, Spain}
%\affiliation{Institute for Theoretical and Computational Physics. Facultad de Ciencias. Universidad de Granada, E-18071, Granada, Spain}

\date{\today }

\begin{abstract}

The kinetic theory description of a low density gas of hard spheres or disks, confined between two parallel plates separated a distance smaller than twice the diameter of the particles, is addressed starting from the Liouville equation of the system.
The associated BBGKY hierarchy of equations for the reduced distribution functions is expanded in powers of a parameter measuring the density of the system in the appropriate dimensionless units. The Boltzmann level of description is obtained by keeping only the two lowest orders in the parameter. In particular, the one-particle distribution function obeys a couple of equations. Contrary to what happens with a Boltzmann-like kinetic equation that has been proposed  for the same system on a heuristic basis, the kinetic theory formulated here admits stationary solutions that are consistent with equilibrium statistical mechanics, both in absence and presence of  external fields.  In the latter case, the density profile is rather complex due to the coupling between the inhomogeneities generated by the confinement and by the external fields. The general theory formulated provides a solid basis for  the study of the properties of strongly confined dilute gases.

\end{abstract}

\maketitle

\section{Introduction}
\label{s1}
Equilibrium and non-equilibrium properties of systems under strongly confined geometries can be very different from that observed in bulk systems. Confinement generates asymmetry and inhomogeneity in the system, and introduces the necessity of modifying the ordinary bulk thermodynamic limit and the consequent simplifications \cite{Ba77}.  In particular, gas transport under confinement exhibits a rich phenomenology and plays a significant role in many areas ranging from energy problems to medical physiology. 
Gas channels regulate the transport across cell membranes and, therefore, confinement is relevant in many biological processes within cellular systems \cite{DZYZyW24}. A comprehensive understanding of the transport properties of gases under strong confinement is not only a requirement for the description of many cellular activities, but it can also inspire nanofluidic devices. The emergence  of  new materials and  the development of novel fabrication techniques of nanodevices have  highlighted peculiar gas transport phenomena in strongly confined systems \cite{SHyR08,ByC10,KNyB21}.

Gas transport can occur in different regimes, depending on the relation between  the size and shape of the system and the mean free path of the particles of the gas. Different regimes require different level of description, ranging from a macroscopic hydrodynamic description governed by particle interactions, to a flow regime in which collisions between gas particles and the the walls dominate, and a macroscopic description in terms of a few fields is not accurate. All these regimes have been extensively studied in bulk gases, and  techniques  have been developed to describe their behavior  in each of the regimes. Nevertheless, the results cannot be directly translated to the case of strongly confined gases. For instance, there is no reason to expect that under strong confinement in one direction, the ordinary Navier-Stokes equations of hydrodynamics still hold in the plane perpendicular to the confinement. In the absence of experimental results leading to phenomenological laws, it is necessary to resort to fundamental descriptions, at the particle level, to try to characterize the different regimes and identify which is the description appropriate for each of them.

Kinetic theory and non-equilibrium statistical mechanics provide the appropriate context to address the above questions. Actually, for dilute bulk gases, the analysis of the kinetic Boltzmann equation has led to a deep understanding of the different gas regimes, and  to identify the relevant variables for the description of the gas in each of them and the evolution equations for these variables \cite{Ce75,FyK72}. Moreover, a powerful numerical technique, named direct Monte Carlo simulation method (DSMC),  has been developed to construct solutions of the Boltzmann equation under extreme boundary conditions \cite{Bi94}. A similar programe was recently started for strongly confined gases. A modified Boltzmann kinetic equation was proposed for these systems \cite{BMyG16,BGyM17,MBGyM22} on the basis of heuristic arguments similar to those used  in the original derivation of the Boltzmann equation for bulk gases. Although formulated for a system of hard spheres or disks strongly confined between two parallel hard walls, the extension to other geometries seems, in principle, straightforward. When  the kinetic equation is  particularized  for an equilibrium situation, results consistent with predictions derived by means of density functional theory \cite{SyL96,SyL97}, and also with molecular dynamics simulations data are found. Also,  agreement with molecular dynamics simulations was found for confined systems composed of hard particles colliding inelastically, that are by definition out of equilibrium  \cite{MGyB19,MGyB19a,MGyB22}.

In spite of  the above success of the proposed kinetic equation, several conflicting points have showed up. The equilibrium density profile obtained from the Boltzmann equation involves all powers of the density, since it is given by an exponential of the density function. This indicates that somehow the modified  Boltzmann equation does not correspond to a well defined low density limit. Although this may seem a merely formal issue, it is quite important to know the range of applicability of the kinetic equation. Even more compellingly is the argument that the modified Boltzmann equation formulated in \cite{BMyG16,MBGyM22} does not lead to the correct equilibrium distribution for hard-sphere or hard-disk systems in presence of a stationary external potential, i. e. the equation does not admit a stationary solution that has the form of the number density times the normalized Mawellian velocity distribution. Quite interestingly, the same happens with the original formulation of the Enskog equation for hard spheres \cite{CyC52} and this was one of the reasons \cite{vB83} validating  the revised Enskog theory formulated by van Beijeren and Ernst \cite{vByE73,note}. The above features indicate the need for a derivation of the kinetic theory description for a low density strongly confined gas of hard spheres or disks starting from the Liouville equation, and considering a mathematically well defined low density limit, as it has been done for the Boltzmann equation for a bulk system \cite{Gr49,La75,DvByK21}. Carrying out such a study is the primary aim of the work being presented here.

The remaining of this paper is organized as follows. In the next section, the geometry of the system to be considered is specified. It is a system of hard spheres or disks confined between two parallel  hard plates separated a distance smaller than twice a  particle diameter. Next, the Liouville equation of the system is formulated, paying special attention to the contribution coming from the confining walls. Its derivation is just a slight modification of that for the bulk system case. The main difference resides in the dominant role played by the characteristic function associated to the boundaries, and defining the space region occupied by the system. This restricts the possible values of the scattering angle between colliding particles and implies the existence of inhomogeneities in the direction perpendicular to the walls.  From the Liouville equation,  the usual Bogoliubov, Born, Green, Kirkwood, and Yvon (BBGKY) hierarchy for reduced distribution functions \cite{Ba77,DvByK21} follows by partial integration. It has the same formal structure as for bulk systems.

In Sec. \ref{s3}, the above hierarchy is written in dimensionless form by scaling space and time. Here is where the confinement plays a crucial role in the analysis. While distances in the direction parallel to the confining plates are scaled with the mean free path of the gas, lengths in the direction perpendicular to them are reduced with the diameter of the particles. Then, a thermodynamic  limit appropriate to the geometry of the system is introduced, and an intensive behavior of the  reduced distribution functions in that limit is assumed \cite{Ba77}. The structure of the dimensionless BBGKY hierarchy suggests the search of solutions in the form of a series expansions in powers of  a parameter  characterizing the density of the system. The details of these calculations are provided in Appendices \ref{ap1} and \ref{ap2}, while only the results to the two lowest orders are given in the main text. The structure of the obtained kinetic theory is different from the usual Boltzmann equation for bulk systems and also from the modified equation in refs. \cite{BMyG16} and \cite{MBGyM22}. The latter is given in Appendix \ref{ap3} for the sake of completeness and to render simpler the comparison with the present results. As a first relevant test of the obtained kinetic theory, in Sec. \ref{s4} it is investigated whether it admits a stationary solution that is consistent with equilibrium statistical mechanics. Systems without and with an external field are considered, and satisfactory answers are obtained in both cases. Finally, Sect. \ref{s5} contains a summary of the conclusions and some final comments and perspectives.

\section{Liouville equation and the BBGKY hierarchy}
\label{s2}
 In this section, the (pseudo)-Liouville equation describing the dynamics of a system of hard spheres or disks under strong confinement will be formulated, 
 as well as the associated BBGKY hierarchy for reduced distribution functions. The system being considered is composed of $N$ elastic hard
 spheres ($d=3$) or disks ($d=2$) of mass $m$ and diameter $\sigma$, confined between two parallel plates ($d=3$) or straight hard walls ($d=2$), separated a distance 
 $h$, that takes values in the interval $\sigma < h < 2 \sigma$. Therefore, particles can not jump one over another in the direction perpendicular to the hard boundaries,
 that is taken as the $z$ axis, so that the walls are located at $z=0$ and $z=h$, respectively. To be specific, the plates are considered to be squares of side $L$ in the three-dimensional case, and a segment of length $L$ for a system of hard disks. The value of $L$ is considered very large, formally infinite. Until it be explicitly indicated (in Sec. \ref{s4}), no external forces acting on the system other than those associated with the hard boundaries are considered.
  
 The particles move along straight lines with constant velocity until they encounter another particle or one of the hard walls. At that point, the velocity of the colliding particles change instantaneously, followed for subsequent straight-line motion at the post-collisional velocity. Here attention is limited to elastic particles and elastic walls, although the theory can be easily extended to inelastic particles (grains) \cite{BDyS97}. Then, when two particles with velocities ${\bm v}_{1}$ and ${\bm v}_{2}$ collide, their velocities instantaneously change to new values ${\bm v}^{\prime}_{1}$ and ${\bm v}^{\prime}_{2}$ defined by
 \begin{eqnarray}
 \label{2.1}
 {\bm v}^{\prime}_{1} &=& {\bm v}_{1}-{\bm g}_{12} \cdot \widehat{\bm \sigma} \widehat{\bm \sigma}, \nonumber \\
 {\bm v}^{\prime}_{2} & = &{\bm v}_{2}+{\bm g}_{12} \cdot \widehat{\bm \sigma} \widehat{\bm \sigma},
 \end{eqnarray}
 where ${\bm g}_{12} \equiv {\bm v}_{1}- {\bm v}_{2}$ is the relative velocity of the particles before the collision, and $\widehat{\bm \sigma}$ is the unit vector directed along the line joining the center of the two particles at contact, and pointing towards particle $1$.
 
 Moreover, when a particle collides with one of the confining hard walls, its velocity ${\bm v}$ is instantaneously changed into ${\bm v}^{\prime (w)}$ given by
 \begin{equation}
 \label{2.2}
 {\bm v}^{\prime (w)} = {\bm v}- 2 v_{z} \widehat{\bm e}_{z},
 \end{equation}
 with $\widehat{\bm e}_{z}$ being the unit vector in the positive direction of the $z$ axis. This collision rule holds for both boundaries.
 
 To formulate the non-equilibrium statistical mechanics of the above system it is necessary to identify the generator of its dynamics. Since the velocities change instantaneously, the associated force is singular and Newton's equations cannot be applied in their usual form. Instead a formalism based on the use of binary collision operators is employed \cite{EDHyvL69,RydL77,McL89,DvByK21}. Consider a phase function $G(\Gamma)$, where $\Gamma$ is a point in the phase space of the system, $\Gamma \equiv \left\{ {\bm r}_{i}, {\bm v}_{i}; i=1, \ldots , N \right\}$, with ${\bm r}_{i}$ being the vector position of particle $i$. The value of $G$ at time $t$ assuming that $\Gamma$ represents the initial positions and velocities of the particles at $t=0$, will be denoted by $S_{t}(\Gamma) G(\Gamma)$. The determination of this quantity for all $\Gamma$ would require to specify the dynamics of two hard spheres or disks when they overlap, and also the dynamics of a particle that is totally or partially outside of the system, as delimited by the hard boundaries. Nevertheless, from a physical point of view, it is clear that  those configurations are irrelevant since they cannot be realized in actual experiments. Therefore, attention will be restricted in the following to consider the expression $W(\Gamma) S_{t}(\Gamma) G(\Gamma)$, where $W(\Gamma)$ is a characteristic function defining the physical non-overlapping region inside the system, i.e.
 \begin{equation}
 \label{2.3}
 W(\Gamma)= W^{(w)}(\left\{ {\bm r}_{i} \right \}) W^{(s)} (\left\{ {\bm r}_{i} \right \}),
 \end{equation}
 where 
 \begin{equation}
 \label{2.4}
 W^{(w)}(\left\{ {\bm r}_{i} \right \}) = \prod_{i=1}^{N} W^{(w)}(z_{i}),
 \end{equation}
 \begin{equation}
 \label{2.5}
 W^{(w)}(z) = \theta \left( z-\frac{\sigma}{2} \right) \theta \left( h-\frac{\sigma}{2} - z\right)
 \end{equation}
 defines the region occupied by the system, and
 \begin{equation}
 \label{2.6}
 W^{(s)} (\left\{ {\bm r}_{i} \right \}) = \prod_{1 \leq i<j \leq N} W_{ij}^{(s)}({\bm r}_{i}, {\bm r}_{j} ),
 \end{equation}
 \begin{equation}
 \label{2.7}
 W_{ij}^{(s)}({\bm r}_{i}, {\bm r}_{j} ) = \theta \left( | {\bm r}_{i}- {\bm r}_{j} |-\sigma \right)
  \end{equation}
 characterizes the non-ovelapping configurations. In the above expressions, $\theta$ is the Heaviside step function defined as
 \begin{equation}
 \label{2.8}
 \theta (x) = \left\{ 
 \begin{array}{ll}
  1  & \mbox{if $x\geq 0$} \\
  0  & \mbox{otherwise}.
  \end{array}
  \right.
 \end{equation}

 The formal theory to be formulated in this section is quite general, in the sense that it can be applied to systems with a wider separations of the two confining plates, and even to systems confined by means of walls of different shapes. All that is needed is to modify consistently the definition of the characteristic functions $ W^{(w)}({\bm r})$, defining the space region inside the system.
 In statistical mechanics, the expectation value of a dynamical variable $G$ at time $t$ is given by
 \begin{equation}
 \label{2.9}
 \langle G(t) \rangle \equiv \int d \Gamma \rho (\Gamma;0) S_{t}(\Gamma) G(\Gamma),
 \end{equation}
 where $\rho(\Gamma;0)$ is the probability density for the initial state $\Gamma$. For the reasons mentioned above, the latter must vanish
 for overlapping configurations and also when a particle is not entirely inside the system. In other words, $\rho(\Gamma;0)$ contains an implicit overlap function $W(\Gamma)$, given by Eq.\, (\ref{2.3}), so that Eq. (\ref{2.9}) is equivalent to
 \begin{equation}
 \label{2.10}
 \langle G(t) \rangle = \int d \Gamma\,  \rho (\Gamma;0) W(\Gamma) S_{t}(\Gamma) G(\Gamma).
 \end{equation}
 Since the dynamics of a hard-sphere or hard-disk system reduces to a succession of instantaneous two particle collisions and of collisions of a particle with the hard walls, the effect 
 of the collisions is additive, and each of them can be studied separately. The calculations have been discussed in detail in \cite{DvByK21}. Although a bulk (not strongly confined system) is considered in this reference, the reasonings remain valid and they will not be reproduced here. It is found that
 \begin{equation}
 \label{2.11}
 W(\Gamma) S_{t}(\Gamma) = W(\Gamma) e^{t \mathcal{L}(\Gamma)},
 \end{equation}
 with the  generator of the dynamics $\mathcal{L}(\Gamma)$  given by
 \begin{equation}
 \label{2.12}
 \mathcal{L}(\Gamma)= \sum_{i=1}^{N} {\bm v}_{i}\cdot {\bm \nabla}_{i} + \sum_{i=1}^{N} \left[  T_{0}^{(w)} (i) + T_{h}^{(w)}(i) \right] + \sum_{1 \leq i<j\leq N} T(i,j).
 \end{equation}
 The first term on the right hand side generates the free streaming, the second term describes the collisions with the confining walls, and the last term takes into account the velocity changes due to particle collisions. The
 walls collision operators are \cite{DyvB77}
 \begin{equation}
 \label{2.13}
 T_{0}^{(w)}(i) = - \theta (-v_{iz}) v_{iz} \delta \left( z_{i}-\frac{\sigma}{2}\right) (b_{i}^{(w)}-1),
 \end{equation}
 \begin{equation}
 \label{2.14}
 T_{h}^{(w)}(i) = \theta (v_{iz}) v_{iz} \delta \left(z_{i}- h + \frac{\sigma}{2}\right) (b_{i}^{(w)}-1).
 \end{equation}
 Here $\delta(x)$ is the Dirac delta function and $b_{i}^{(w)}$ is an operator changing the velocities ${\bm v}_{i}$ to its right into ${\bm v}_{i}^{\prime (w)}$, accordingly with the collision rule given in Eq. (\ref{2.2}). These expressions describe the conservation of the flux of particles that collide with the walls.
The binary collision operator $T(i,j)$ for particles $i$ and $j$ is
\begin{equation}
\label{2.15}
T(i,j)= \sigma^{d-1} \int d \widehat{\bm \sigma}\,  \theta (-{\bm g}_{ij} \cdot \widehat{\bm \sigma}) | {\bm g}_{ij} \cdot \widehat{\bm \sigma} | \delta({\bm r}_{ij}-{\bm \sigma}) \left[ b_{\bm \sigma} (i,j)-1 \right],
\end{equation}
where $ d \widehat{\bm \sigma}$ denotes the solid angle element for the unit vector $\widehat{\bm \sigma}$, ${\bm \sigma}= \sigma \widehat{\bm \sigma}$, ${\bm r}_{ij} \equiv {\bm r}_{i}-{\bm r}_{j}$ is the relative position vector of the two particles, and $b_{\bm \sigma} (i,j)$ is an operator which replaces all the velocities ${\bm v}_{i}$  and ${\bm v}_{j}$ appearing to its right by the post-collisional velocities ${\bm v}^{\prime}_{i}$ and ${\bm v}^{\prime}_{j}$ given by Eqs. (\ref{2.1}).

Equation (\ref{2.10}) can be expressed in the equivalent form 
\begin{equation}
\label{2.16}
 \langle G(t) \rangle = \int d \Gamma  \left[ \overline{S}_{t}(\Gamma) W(\Gamma)  \rho (\Gamma;0)  \right]  G(\Gamma)\, 
  = \int d\Gamma\, \rho (\Gamma,t) G(\Gamma),
\end{equation}
with the definition
\begin{equation}
\label{2.17}
\rho(\Gamma,t) \equiv \overline{S}_{t}(\Gamma)  W(\Gamma) \rho (\Gamma;0).
\end{equation}
The adjoint time evolution operator $\overline{S} _{t}(\Gamma)$ verifies
\begin{equation}
\label{2.18}
\overline{S}_{t}(\Gamma) W(\Gamma)= e^{t \overline{\mathcal{L}}} W(\Gamma)
\end{equation}
where the generator of the dynamics is now
\begin{equation}
\label{2.19}
\overline{\mathcal L} (\Gamma)= -\sum_{i=1}^{N} {\bm v}_{i}\cdot {\bm \nabla}_{i} + \sum_{i=1}^{N} \left[  \overline{T}_{0}^{(w)} (i) + \overline{T}_{h}^{(w)}(i) \right] + \sum_{1 \leq i<j\leq N} \overline{T}(i,j).
\end{equation}
The new walls and binary collision operators are defined as
\begin{equation}
\label{2.20}
\overline{T}_{0}^{(w)}(i) = v_{iz} \delta \left( z_{i}-\frac{\sigma}{2} \right) \left[ \theta (v_{iz}) b_{i}^{(w)}+\theta (-v_{iz}) \right],
\end{equation}
\begin{equation}
\label{2.21}
\overline{T}_{h}^{(w)}(i) = -v_{iz} \delta \left( z_{i} - h +\frac{\sigma}{2} \right) \left[ \theta (-v_{iz}) b_{i}^{(w)}+\theta (v_{iz}) \right],
\end{equation}
\begin{equation}
\label{2.22}
\overline{T}(i,j) = \sigma^{d-1} \int d \widehat{\bm \sigma}\,  \theta ({\bm g}_{ij} \cdot \widehat{\bm \sigma}) | {\bm g}_{ij} \cdot \widehat{\bm \sigma} | \left[  \delta({\bm r}_{ij}-{\bm \sigma}) b_{\bm \sigma} (i,j)-\delta ({\bm r}_{ij}+ {\bm \sigma}) \right].
\end{equation}
A physically important property of the evolution operators is that they conserve the non-overlaping conditions and keeps all the particles inside of the system, i.e.
\begin{equation}
\label{2.23}
\overline{S}_{t}(\Gamma) W(\Gamma)= W(\Gamma) \overline{S}_{t} (\Gamma) W(\Gamma)
\end{equation}
and, therefore,
\begin{equation}
\label{2.24}
\rho(\Gamma,t) = W ( \Gamma) \rho (\Gamma,t).
\end{equation}
This property reflects that nonphysical configurations cannot be generated by the dynamics of the system when starting from physical configurations. The $N$-particle distribution $\rho (\Gamma,t)$ evolves in time according to the pseudo-Liouville equation,
\begin{equation}
\label{2.25}
\left ( \frac{\partial}{\partial t}\, - \overline{\mathcal L} \right) \rho (\Gamma,t) =0.
\end{equation}
Reduced  $s$-particle distribution functions are defined as
\begin{equation}
\label{2.26}
f_{s}(x_{1},\ldots,x_{s},t) = \frac{N!}{(N-s)!} \int dx_{s+1} \cdots dx_{N} \rho(\Gamma,t),
\end{equation}
where $x_{i}$ denotes the position and velocity of particle $i$ and $dx_{i} \equiv d{\bm r}_{i} d{\bm v}_{i}$. These functions obey the BBGKY hierarchy that follows in the usual way by partial integration of the pseudo-Liouville equation over $x_{s+1}, \ldots,x_{N}$ \cite{Ba77,DvByK21,RydL77},
\begin{equation}
\label{2.27}
\left[ \frac{\partial}{\partial t} - \overline{L} ( x_{1}, \ldots,x_{s}) \right] f_{s}(x_{1},\ldots,x_{s},t) = \sum_{i=1}^{s} \int dx_{s+1}\, \overline{T} (i,s+1) f_{s+1}(x_{1},\ldots,x_{s+1},t).
\end{equation}
In this expression,  $\overline{L} ( x_{1}, \ldots,x_{s})$ is the generator for the dynamics of the distribution function as defined in Eq. (\ref{2.19}), but for a system of $s$ particles.

 The above hierarchy is formally the same as for a bulk, not strongly confined system \cite{DvByK21}.  Actually, it is also valid for other kinds of walls, such as frictional walls, diffusive walls,  or walls moving in the horizontal direction \cite{DyvB77}. From Eq.\ (\ref{2.24}) and the definition (\ref{2.26}) it follows that
\begin{equation}
\label{2.28}
f_{s}(x_{1},\ldots,x_{s},t) = W (\Gamma_{s}) f_{s}(x_{1},\ldots,x_{s},t),
\end{equation}
with $W(\Gamma_{s})$ being the same characteristic function as defined in Eq. (\ref{2.3}), but restricted to the configurational space of $s$ particles. It will be shown in the following that this has very relevant consequences when studying the properties of  the hierarchy in some specific limits.

\section{Boltzmann theory for the strongly confined gas}
\label{s3}
A dimensionless form of the BBGKY hierarchy, Eq.\ (\ref{2.27}), is obtained by scaling the space and time with the appropriate  scales \cite{DLyB95}. The latter are suggested by the expected behavior of the reduced distribution functions in the limit of a large system. Taking into account the confinement imposed by the two hard boundaries, it will be assumed that in the limit $L \rightarrow \infty$, $N \rightarrow \infty$, while $\tilde{n} \equiv N/L^{d-1}$ is kept  finite, the reduced distribution functions $f_{s}$ becomes intensive properties, i.e. independent of $L$ and $N$ separately \cite{Ba77}.  Consistently with this idea, the spatial coordinates parallel to the confining walls are scaled with $\ell_{0} \equiv (\tilde{n} \sigma^{d-2})^{-1}$, that is a quantity of the order of the mean free path associated to the motion parallel to the walls. On the other hand, the z coordinates are scaled with $\sigma$, that is a measure of the range of the inter-particle force and also of the separation between the two confining walls. Moreover, the time will be scaled relative to the mean free time, $ \tau_{0} \equiv \ell_{0}/v_{0}$, $v_{0}$ being some characteristic velocity of the particles. The reduced distribution functions $f_{s}$ are scaled with $\left[ v_{0}^{d}/(\tilde{n} \sigma^{-1}) \right]^{s}$, so that the scaled distributions also verify the hypothesis made above. Then, the resulting dimensionless hierarchy contains the parameter $\alpha \equiv \sigma/\ell_{0}  = \widetilde{n} \sigma^{d-1}$, the ratio of the force range to the mean free path in the plane. This parameter is small at low density, suggesting an expansion of the distribution functions in powers of $\alpha$. Details of the calculation are provided in Appendix \ref{ap1}, while here only the results for the two lowest orders in $\alpha$  are given.
The one-particle distribution function has the form
\begin{equation}
\label{3.1}
f_{1}(x_{1},t)= f_{1}^{(0)}(x_{1},t) + \alpha f_{1}^{(1)}(x_{1},t).
\end{equation}
To formulate in a precise way the equations obeyed by $f_{1}^{(0)}$ and $f_{1}^{(1)}$, it is convenient to introduce functions
$\widetilde{f}_{1}^{(0)}$ and $\widetilde{f}_{1}^{(1)}$ by means of the relations
\begin{equation}
\label{3.2}
f_{1}^{(0)}(x,t) = W^{(w)}(z) \widetilde{f}_{1}^{(0)}(x,t), \quad f_{1}^{(1)}(x,t) = W^{(w)}(z) \widetilde{f}_{1}^{(1)}(x,t).
\end{equation}
These functions can be assumed to be continuous at the hard boundaries, and are found to be solutions to the equations
\begin{equation}
\label{3.3}
\frac{\partial}{\partial z _{1}} \widetilde{f}_{1}^{(0)}(x_{1},t) =0,
\end{equation}and
\begin{eqnarray}
\label{3.4}
 \lefteqn{W^{(w)} (z_{1}) \left[ \left( \frac{\partial}{\partial t}  +   {\bm v}_{1\perp} \cdot {\bm \nabla}_{1 \perp} \right) \widetilde{f}_{1}^{(0)}(x_{1},t)  +\alpha  v_{1z}\frac{\partial}{\partial z_{1}} \widetilde{f}_{1}^{(1)}(x_{1},t)\right] }  \nonumber \\
&& =   W^{(w)} (z_{1}) \int dx_{2}\,  \overline{T}_{0} (1,2) \widetilde{f}_{1}^{(0)}(x_{1},t)  W^{(w)} (z _{2}) \widetilde{f}_{1}^{(0)}(x_{2},t).
\end{eqnarray}
Here $\overline{T}_{0}(1,2) $ is the modified binary collision operator for hard spheres,
\begin{equation}
\label{3.5}
\overline{T}_{0}(1,2) \equiv  \sigma^{d-1}  \delta({\bm r}_{12 \perp}) \int d \widehat{\bm \sigma}\,  \theta ({\bm g}_{12} \cdot \widehat{\bm \sigma}) | {\bm g}_{12} \cdot \widehat{\bm \sigma} | \left[  \delta (z_{12} -\sigma_{z}) b_{\bm \sigma} (1,2)- \delta (z_{12}+\sigma_{z})  \right],
\end{equation}
and the subindex $\perp$ is used to denote the projection of a vector on the plane parallel to the hard walls.  The effect of the strong confinement manifests itself in the presence of the characteristic function $W^{(w)}(z)$ in the collision term on the right-hand-side of Eq. (\ref{3.4}). Had we considered a system with a wider separation of the walls, the above result would remain valid, if the characteristic function is substituted by the one adapted to the new geometry.
Moreover, $\widetilde{f}_{1}^{(0)}(x,t)$ must be an even function of $v_{z}$ and $\widetilde{f}_{1}^{(1)}(x,t) $ must verify the boundary conditions
\begin{equation}
\label{3.6}
\theta(v_{z}) \delta \left(z- \frac{\sigma}{2} \right) \widetilde{f}_{1}^{(1)}(x,t) =\theta(v_{z}) \delta \left(z- \frac{\sigma}{2} \right) b^{(w)} \widetilde{f}_{1}^{(1)}(x,t), 
\end{equation}
 \begin{equation}
 \label{3.7}
 \theta(-v_{z}) \delta \left(z- h + \frac{\sigma}{2} \right) \widetilde{f}_{1}^{(1)}(x,t) =\theta(- v_{z}) \delta \left(z- h+\frac{\sigma}{2} \right) b^{(w)} \widetilde{f}_{1}^{(1)}(x,t).
 \end{equation}
 Equations (\ref{3.3}) and (\ref{3.4}) correspond to  the Boltzmann level of description in the limit of a strongly confined system of hard spheres or disks 
 as considered here. Moreover, a two-particle correlation function $g_{2}(x_{1},x_{2},t)$ can be defined by 
 \begin{equation}
 \label{3.8}
 g_{2}(x_{1},x_{2},t) \equiv f_{2}(x_{1},x_{2},t) -f_{1}(x_{1},t) f_{1}(x_{2},t).
 \end{equation}
 The expansion of $f_{1}$ and $f_{2}$ in powers of $\alpha$ generates a similar expansion for $g_{2}$. To lowest order, it is
 \begin{equation}
 \label{3.9}
 g_{2}^{(0)} (x_{1},x_{2},t)=0,
 \end{equation}
 for both $d=2$ and $d=3$. Nevertheless, to the next order
 \begin{equation}
 \label{3.9a}
 g_{2}^{(1)} (x_{1},x_{2},t) \equiv f_{2}^{(1)} (x_{1},x_{2},t) -f_{1}^{(0)} (x_{1},t) f_{1}^{(1)} (x_{2},t)- f_{1}^{(0)} (x_{2},t) f_{1}^{(1)} (x_{1},t),
\end{equation}
 vanishes for $d=3$, but for $d=2$ it  does not,  being given by a solution of the equation
 \begin{eqnarray}
 \label{3.10}
 \lefteqn{W^{(w)} (z_{1}) W^{(w)} (z_{2}) \alpha \left( v_{1z} \frac{\partial}{\partial z_{1}} + v_{2z} \frac{\partial}{\partial z_{2}} \right) \widetilde{g}_{2}^{(1)}(x_{1},x_{2},t)}
 \nonumber \\
 & &= W^{(w)} (z_{1}) W^{(w)} (z_{2})\overline{T}_{0}(x_{1},x_{2}) f_{1}^{(0)} (x_{1},t) f_{2}^{(0)} (x_{2},t)
 \end{eqnarray}
 with the boundary conditions
 \begin{equation}
 \label{3.11}
 \theta (v_{1z}) \delta \left( z_{1}- \frac{1}{2} \right) W^{(w)}(z_{2}) \widetilde{g}_{2}^{(1)}(x_{1},x_{2},t)
= \theta(v_{1z})  \delta \left( z_{1}- \frac{1}{2} \right) W^{(w)}(z_{2}) b_{1}^{(w)} \widetilde{g}_{2}^{(1)}(x_{1},x_{2},t),
\end{equation}
\begin{equation}
 \label{3.12}
 \theta (-v_{1z}) \delta \left( z_{1}- h + \frac{\sigma}{2} \right) W^{(w)}(z_{2}) \widetilde{g}_{2}^{(1)}(x_{1},x_{2},t)
= \theta(v_{1z})  \delta \left( z_{1}- h + \frac{\sigma}{2} \right) W^{(w)}(z_{2}) b_{1}^{(w)} \widetilde{g}_{2}^{(1)}(x_{1},x_{2},t),
\end{equation}
and the two equalities following by interchanging the subindices 1 and 2 everywhere  in the two equations. In Eqs. (\ref{3.10})-(\ref{3.12}), the function $\widetilde{g}_{2}$ is related with the correlation function $g_{2}$ by
\begin{equation}
\label{3.13}
g_{2}(x_{1},x_{2},t) \equiv   W^{(w)} (z_{1}) W^{(w)} (z_{2})   \widetilde{g}_{2}(x_{1},x_{2},t).
\end{equation}
Of course, the function $g_{2}(x_{1},x_{2},t)$ also contains an implicit $W^{(s)} ({\bm r}_{1}, {\bm r}_{2})$ factor, that has relevant consequences at moderate values of the density, i.e.  for values of $\alpha$ larger than those being addressed here \cite{Lu96}.

The expansion method also provides the form of all the the distribution function functions $f_{s}^{(0)
}$ and $f_{s}^{(1)}$, for $s\geq 3$. Their explicit forms are given in Eqs. (\ref{a12}), (\ref{a20}), and (\ref{a26}). 

To render simpler the comparison of the above low density theory for a strongly confined gas with the usual Boltzmann equation and also for applications of the theory, it is worth to rewrite Eq. (\ref{3.4})  in a form looking closer to the Boltzmann equation:
\begin{equation}
\label{3.14}
 W^{(w)} (z_{1}) \left[ \left( \frac{\partial}{\partial t}  +   {\bm v}_{1\perp} \cdot {\bm \nabla}_{1 \perp} \right) \widetilde{f}_{1}^{(0)}(x_{1},t)  +\alpha  v_{1z}\frac{\partial}{\partial z_{1}} \widetilde{f}_{1}^{(1)}(x_{1},t)\right]   = W^{(w)}(z_{1})J[x_{1}|\widetilde{f}_{1}^{(0)}].
\end{equation}
For a system of hard spheres ($d=3$), the collision term is given by
\begin{eqnarray}
\label{3.15}
J[x_{1}|f]  &= & \sigma \int_{\sigma/2}^{h- \sigma/2} dz_{2} \int d{\bm v}_{2} \int_{0}^{2 \pi} d \phi\, | {\bm g}_{12} \cdot \widehat{\bm \sigma} |   \left[ \theta ( {\bm g}_{12} \cdot \widehat{\bm \sigma} )  b_{\bm \sigma} (12) - \theta ( -{\bm g}_{12} \cdot \widehat{\bm \sigma}) \right]  \nonumber \\
&& \times f({\bm r}_{1},{\bm v}_{1} ,t) f({\bm r}_{1},{\bm v}_{2},t).
\end{eqnarray}
In this expression, it is understood that $\widehat{\bm \sigma}$ has the components
\begin{equation}
\label{3.16}
\widehat{\bm \sigma} \equiv \left \{ \sin \theta \sin \phi, \sin \theta \cos \phi, \cos \theta \right\}
\end{equation}
with
\begin{equation}
\label{3.17}
\cos \theta =\frac{z_{12}}{\sigma}\, , \quad \sin \theta = + \sqrt {1- \left( \frac{z_{12}}{\sigma} \right)^{2} }.
\end{equation}
In the case of a system of confined hard disks ($d=2$), the collision term  is
\begin{eqnarray}
\label{3.18}
J[x_{1}|f] & = & \int_{\sigma/2}^{h- \sigma/2} dz_{2} \int d{\bm v}_{2} \frac{1}{\cos \theta_{1}}\, \left( 1 + \mathcal{P}_{{\bm \sigma}_{1}, {\bm \sigma}_{2}} \right) | {\bm g}_{12} \cdot \widehat{\bm \sigma}_{1} |   
\nonumber \\
&& \times\left[ \theta ( {\bm g}_{12} \cdot \widehat{\bm \sigma}_{1} )  b_{\bm \sigma} (12) - \theta ( -{\bm g}_{12} \cdot \widehat{\bm \sigma}_{1}) \right]  
 f({\bm r}_{1},{\bm v}_{1} ,t) f({\bm r}_{1},{\bm v}_{2},t).
 \end{eqnarray}
Here $ \mathcal{P}_{{\bm \sigma}_{1}, {\bm \sigma}_{2}}$ is an operator permuting the vectors $\widehat{\bm \sigma}_{1}$ and $\widehat{\bm \sigma}_{2}$ to its right. These vectors have components
\begin{equation}
\label{3.19}
\widehat{\bm \sigma}_{i} \equiv  \left\{ \sin \theta , \cos \theta_{i} \right\},
\end{equation}
\begin{equation}
\label{3.20}
\sin \theta = \frac{z_{12}}{\sigma}, \quad \cos \theta_{i} = (-1)^{i+1} \sqrt {1- \left( \frac{z_{12}}{\sigma} \right)^{2} },
\end{equation}
$i=1,2$. A sketch of the derivation of Eqs. (\ref{3.15}) and (\ref{3.18}) is presented in   Appendix \ref{ap2}. 

The structure of Eqs.\ (\ref{3.3}) and (\ref{3.4}) is quite peculiar. The former only establishes that to zeroth order the one-particle distribution function, $f_{1}^{(0)}$, is homogeneous along the direction perpendicular to the plates, while the latter involves this distribution function through its temporal evolution, its gradients in the plane parallel to the plates, and the effect of collisions. Nevertheless, Eq. (\ref{3.4}) is not closed for $f_{1}^{(0)}$ and contains $f_{1}^{(1)}$ through its gradient in the direction perpendicular to the plates. In spite of this apparent complexity, it will be shown in the following sections that the description given by Eqs.\ (\ref{3.3}) and (\ref{3.4}) allows the explicit calculations of  properties of the system under quite physically relevant conditions.

As already mentioned in the Introduction section, a modified Boltzmann kinetic equation for strongly confined systems of dilute hard spheres and disks, with the same geometry as considered here, has been previously proposed on the basis of heuristic arguments, similar to those employed to derive the original Boltzmann equation for bulk systems \cite{BMyG16,BGyM17,MBGyM22}. Contrary to the  present theory, it is a closed kinetic equation for the one-particle distribution function, expected to be valid in the low density limit, similarly to the standard Boltzmann equation. 
For completeness, the modified Boltzmann  equation is  reproduced in Appendix \ref{ap3} in terms of the notation employed in this paper, both for $d=2$ and $d=3$, so that  it can be easily compared with the present theory. A main important difference is that the collision term involves one-particle distribution functions with possible dependence on the $z$-coordinates, and takes into account that two colliding particles have, in general, different values of this vertical coordinate.

For the sake of simplicity in the notation, the tilde on the distribution functions and (sometimes) the characteristic functions $W^{(w)}$ and $W^{(s)}$ will be omitted from now on, but  it is quite important to keep them in mind when applying the developed  kinetic theory.

\section{The equilibrium density profile}
\label{s4}
\subsection{In absence of external fields}
Suppose that the strongly confined gas is at equilibrium with temperature $T$. Because of the symmetry of the system, the one-particle distribution function is expected to have the form
\begin{equation}
\label{4.1} f_{1,eq}({\bm r},{\bm v},t) = n(z) \varphi ({\bm v}),
\end{equation}
where $ k_{B}$ is the Boltzmann constant, $\varphi ({\bm v})$ is the Maxwellian
\begin{equation}
\label{4.2}
\varphi({\bm v}) = \left( \frac{m}{2 \pi k_{B} T} \right)^{d/2} e^{-\frac{mv^{2}}{2k_{B} T}},
\end{equation}
and $n(z)$ is the number of particles density that depends only on $z$, being symmetric around $z=h/2$. The $\alpha$-expansion introduced in the previous section, when applied to the distribution $f_{1,eq}$ given above, keeping only up to first order reads
\begin{equation}
\label{4.3}
f_{1,eq}({\bm r},{\bm v}) = f_{1,eq}^{(0)} ({\bm v}) + \alpha f_{1,eq}^{(1)} (z,{\bm v}),
\end{equation}
with
\begin{equation}
\label{4.4}
f_{1,eq}^{(0)} ({\bm v})= n^{(0)} \varphi ({\bm v})
\end{equation}
and
\begin{equation}
\label{4.5}
f_{1,eq}^{(1)} ({\bm v})= \frac{n^{(1)}(z)}{\alpha}\, \varphi ({\bm v}).
\end{equation}
The homogeneous number density $n^{(0)}$ is
\begin{equation}
\label{4.6}
n^{(0)} \equiv \frac{N}{L^{d-1} (h-\sigma)}\, .
\end{equation}
Of course, the distribution function given in Eq. (\ref{4.4}) satisfies  the two conditions found in the previous section: it is homogeneous along $z$ and it is an even function of $v_{z}$. Also, the expression of $f_{1,eq}^{(1)}$ in Eq. (\ref{4.5}) accomplishes the boundary conditions (\ref{3.6}) and (\ref{3.7}).

In the following, the cases of hard spheres ($d=3$) and hard disks ($d=2$) will be considered separately. For hard spheres, substitution of Eqs. (\ref{4.4}) and (\ref{4.5}) into Eq. (\ref{3.4}) after some algebra leads to
\begin{equation}
\label{4.7}
 \frac{\partial}{\partial z}  n^{(1)} (z)  = \pi  n^{(0)2} (h-\sigma)  \left( 2z - h \right).
\end{equation}
The fact that the right-hand-side of this equation is different from zero is  because the collision term in Eq.\ (\ref{3.4}) does not vanish for Maxwellian velocity distributions.  Eq. (\ref{4.7})  can be easily integrated to give
\begin{equation}
\label{4.8}
 n^{(1)} (z) = \frac{ \pi n^{(0)2} (h -\sigma)}{4}  \left[ (2z-h)^{2} - \frac{1}{3} (h-\sigma)^{2} \right] 
\end{equation}
where the normalization condition
\begin{equation}
\label{4.9}
\int_{\sigma/2}^{h- \sigma/2} dz\,  n^{(1)}(z) =0
\end{equation}
have been employed. Therefore, up to first order in $\alpha$ the equilibrium density profile is
\begin{equation}
\label{4.10}
n(z) \approx n^{(0)} + n^{(1)}(z) = n^{(0)} \left\{ 1+ \frac{ \pi n^{(0)} (h -\sigma)}{4}  \left[ (2z-h)^{2} - \frac{1}{3} (h-\sigma)^{2} \right] \right\}.
\end{equation}
Of course, this expression holds for $\sigma/2 \leq z \leq h- \sigma/2$.
For hard disks, Eqs. (\ref{3.4}), (\ref{4.4}) and (\ref{4.5})  lead to the differential equation
\begin{equation}
\label{4.10a}
\frac{\partial}{\partial z_{1}}  n^{(1)} (z_{1}) = \frac{2 n^{(0)} \alpha}{\sigma (h-\sigma)} \int_{\sigma /2}^{h-\sigma/2} dz_{2}\, \frac{z_{2}}{ \sqrt{\sigma^{2}- z_{12}^{2}}}.
\end{equation}
Although the integral on the right-hand-side is trivial, the posterior integral over $z_{1}$, implying the presence of an arbitrary constant to be determined by normalization, leads to a rather complicated expression that is not very illuminating. A useful approximation can be found by considering the limit of a very narrow channel, i. e. $h-\sigma \ll \sigma$. In this limit, the allowed values of $z_{12}$ inside the integral in Eq. (\ref{4.10}) verify $|z_{12}| \ll \sigma$ and to lowest order $\sqrt{\sigma^{2}- z_{12}^{2}} \approx \sigma$. Then, the solution of the equation is easily obtained and, after normalization, it  is found that the density profile inside the system is 
\begin{equation}
\label{4.11}
n(z) = n^{(0)} \left\{ 1 + \frac{   n^{(0)} (h -\sigma)}{4 \sigma }  \left[ (2z-h)^{2} - \frac{1}{3} (h-\sigma)^{2} \right] \right\}.
\end{equation}

This expression is very similar to Eq. (\ref{4.10}) for a system of hard spheres. However, while the latter is an exact consequence of the kinetic equation (\ref{3.4}) and valid for any separation of the plates between $\sigma$ and $ 2 \sigma$, the profile given by Eq.\ (\ref{4.11}) has been obtained in the limit  $h \rightarrow \sigma^{+}$. Note that these calculations confirm the guess made for the form of the steady solution of the kinetic theory developed in the above sections for the one-particle distribution function of a strongly confined system of hard spheres or disks in absence of external fields.. 

The stationary density profile for a system of hard spheres given in Eq. (\ref{4.10}) agrees with the low density limit (small $n^{(0)}$ keeping $h$ constant) of the expression derived by Schmidt and L\"{o}wen \cite{SyL96,SyL97}, using equilibrium density functional theory and also with the low density limit of the expression derived using the Boltzmann-like kinetic equation revised in Appendix \ref{ap3} \cite{BMyG16,BGyM17}. Also, the expression for hard disks, Eq. (\ref{4.11})  coincides with the expansion up to second order in the density of the  profile obtained in Ref. \cite{MBGyM22}, where the limit of a very narrow channel was also considered. Since the accuracy of the density profiles predicted by means of the modified Boltzmann kinetic equation have been checked at low density by mean of molecular dynamics simulations \cite{BMyG16,MBGyM22,BBGyM17}, there is no need to verify the results predicted here; both agree in the low density limit. 
On the other hand, in this context it is worth to remark that the modified Boltzmann equation leads to profiles that depend exponentially on the density, something that  
seems in contradiction with the expectation that  the Boltzmann equation corresponds to the low density limit. Mathematically, this behavior follows from the $z$ dependence of the one-particle distribution functions appearing in the collision term, something that does not happens in the kinetic theory formulated here.

\subsection{In presence of an external field in the horizontal plane}
Consider now that the confined gas is again at equilibrium at temperature $T$, but the particles are  submitted to an external force of the form
\begin{equation}
\label{4.12}
{\bm F}({\bm r}_{\perp}) =- {\bm \nabla}_{\perp} U({\bf r}_{\perp}),
\end{equation}
i.e. the force acts in the plane parallel to the plates. The potential energy function is otherwise arbitrary. Given that the above external force does not depend on $z$, it seems clear that the same scaling expansion as considered in Sec.\ \ref{s3}  leads to equation
\begin{eqnarray}
\label{4.13}
 \lefteqn{\left[ \frac{\partial}{\partial t}  +   {\bm v}_{1\perp} \cdot {\bm \nabla}_{1 \perp}    + \frac{{\bm F} ({\bm r}_{\perp})}{m} \cdot \frac{\partial}{\partial {\bm v}_{\perp}} \right] f_{1}^{(0)}(x_{1},t)+\alpha  v_{1z}\frac{\partial}{\partial z_{1}} f_{1}^{(1)}(x_{1},t)}  \nonumber \\
&& =   \int dx_{2}\,  \overline{T}_{0} (1,2) f_{1}^{(0)}(x_{1},t) f_{1}^{(0)}(x_{2},t).
\end{eqnarray}
On the basis of equilibrium statistical mechanics, we search by a solution of the form given in Eqs. (\ref{4.3})-(\ref{4.5}), but now the zeroth order density profile is
\begin{equation}
\label{4.14}
n^{(0)} ({\bm r}_{\perp}) = C e^{-\frac{U({\bm r}_{\perp})}{k_{B} T}},
\end{equation}
where $C$ is the normalization constant
\begin{equation}
\label{4.15}
C= \frac{N}{(h-\sigma) \int_{L^{d-1}} d{\bm r}_{\perp} e^{- \frac{U({\bm r}_{\perp})}{k_{B}T}}}\, .
\end{equation}
The integral over ${\bm r}_{\perp}$ extends over the horizontal plane or line parallel to the confining plates. The distribution function $f_{1,eq}^{(0)} (x) = n^{(0)} ({\bm r}_{\perp}) \varphi ({\bm v})$ with $n^{(0)}({\bm r}_{\perp})$ given by Eq. (\ref{4.14}) satisfies the equation
\begin{equation}
\label{4.16}
 {\bm v}_{1 \perp} \cdot {\bm \nabla}_{1 \perp} f_{1,eq}^{(0)} (x_{1}) = - \frac{{\bm F} ({\bm r}_{\perp} )}{m} \cdot \frac{ \partial}{\partial {\bm v}_{1 \perp}}
f_{1,eq}^{(0)} (x_{1}).
\end{equation}
Therefore, particularization of Eq.\, (\ref{4.13}) for equilibrium yields
\begin{eqnarray}
\label{4.17}
v_{1z} \frac{\partial }{\partial z_{1}} n^{(1)} ({\bm r}_{1}) &  = & \sigma^{d-1}    \left[ n^{(0)} ({\bm r}_{1 \perp}) \right]^{2} 
 \int dz_{2}  \int d{\bm v}_{2} \int d \widehat{\bm \sigma}\, \theta \left( {\bm g}_{12} \cdot \widehat{\bm \sigma} \right) | {\bm g}_{12} \cdot \widehat{\bm \sigma} |
 \nonumber \\
&& \times \varphi ({\bm v}_{2} )  W^{((w)}(z_{2}) W^{((w)}(z_{2})\left[ \delta ( z_{12}- \sigma_{z}) - \delta (z_{12} + \sigma_{z}) \right]  \nonumber \\
&=& 
 \sigma^{d-1}    \left[ n^{(0)} ({\bm r}_{1 \perp}) \right]^{2} 
 \int dz_{2}  \int d{\bm v}_{2} \int d \widehat{\bm \sigma}\,   {\bm g}_{12} \cdot \widehat{\bm \sigma} 
 \nonumber \\
&& \times \varphi ({\bm v}_{2} ) W^{((w)}(z_{2}) \delta (z_{12}- \sigma_{z})  \nonumber \\
&=& \sigma^{d-1}    \left[ n^{(0)} ({\bm r}_{1 \perp}) \right]^{2}  v_{1z}
 \int dz_{2}   \int d \widehat{\bm \sigma}\,  \delta (z_{12} - \sigma_{z}) \widehat{\sigma}_{z} W^{((w)}(z_{2}).  \nonumber \\
 \end{eqnarray}
Once more time, the cases $d=3$ and $d=2$ are considered separately. For a system of hard spheres, the integrals on the right hand side of the above equation can be easily evaluated, leading to the differential equation
\begin{equation}
\label{4.18}
\frac{\partial }{\partial z} n^{(1)} ({\bm r}) = \pi  \left[ n^{(0)} ({\bm r}_{ \perp}) \right]^{2}  (2z-h) (h-\sigma).
\end{equation}
Now, this equation can be solved and after requiring the normalization condition (\ref{4.9}) it is found
\begin{equation}
\label{4.19}
n^{(1)} ({\bm r}) = \pi (h -\sigma) \left\{ \left[ n^{(0)} ({\bm r}_{ \perp}) \right]^{2}  \left( z- \frac{h}{2} \right)^{2} - \frac{(h-\sigma)^{2}}{12 L^{2}}\, \int_{L^{2}}d{\bm r}_{\perp} \left[ n^{(0)} ({\bm r}_{\perp}) \right]^{2} \right\}.
\end{equation}
 A possible function  of ${\bm r}_{\perp}$ has been omitted, since according with the analysis performed is expected to be of next order in the density ($\alpha$). In summary, the density profile in the low density limit being considered is given by
\begin{equation}
\label{4.20}
n({\bm r}) = n^{(0)} ({\bm r}_{\perp} )+ n^{(1)} ({\bm r}),
\end{equation}
where
\begin{equation}
\label{4.20a}
n^{(0)} ({\bm r}_{\perp})=  \frac{N e^{-\frac{U({\bm r}_{\perp})}{k_{B}T}}}{ (h-\sigma) \int_{L^{2}} d{\bm r}_{\perp} e^{- \frac{U({\bm r}_{\perp})}{k_{B} T}}}
\end{equation}
and the function $n^{(1)}({\bm r})$ is  defined in Eq.\ (\ref{4.19}).

For hard disks ($d=2$), using a  method similar to the one outlined in Appendix \ref{ap2}, it is obtained
\begin{equation}
\label{4.21}
\frac{\partial }{\partial z_{1}} n^{(1)} ({\bm r}_{1}) = \left[ n^{(0)} ( r_{1 \perp}) \right]^{2} \int_{\sigma/2}^{h-\sigma/2} dz_{2}\,  \frac{z_{12}}{\sqrt{1- \left( \frac{z_{12}}{\sigma} \right)^{2} }}\, .
\end{equation}
For the same reason as invoked in absence of external field, the limit of a very narrow channel, $h-\sigma \ll \sigma$, will be considered now. In this approximation, 
the solution of the above equation can be directly obtained and, after requiring the normalization condition (\ref{4.9}) and omitting a function of the coordinate perpendicular to the $x$-axis, the expression for $n^{(1)} ({\bm r}_{1})$ follows.
The result is
\begin{equation}
\label{4.22}
n^{(1)} ({\bm r}) = \frac{h-\sigma}{\sigma} \left\{ \left[ n^{(0)} (r_{\perp}) \right]^{2} \left( z -\frac{h}{2} \right)^{2} - \frac{h-\sigma}{12 L}
\int_{L} d r_{\perp}\, \left[ n^{(0)} (r_{\perp})\right]^{2} \right\}.
\end{equation}
It is worth to mention again that, in spite of the similarity between Eqs. (\ref{4.19}) and (\ref{4.22}), they correspond to different degrees of generality, since the latter is restricted to the case of extreme confinement. Of course, the density field up to first order is given by Eq. (\ref{4.20}), and $n^{(0)}({\bm r}_{\perp})$ by Eq. (\ref{4.20a}), although now the integral extends over the length $L$ of the hard boundaries.

Several comments seem appropriate at this point. Firstly, the above results show the consistency of the assumed form of the steady distribution function in presence of the external field. This is a quite relevant point, since the modified Boltzmann kinetic equation considered in some previous works 	\cite{BMyG16,BGyM17,MBGyM22}, and given in Appendix \ref{ap3}, when including an external force does not admit a stationary solution of the form $f_{1,st}({\bm r},{\bm v}) = n({\bm r}) \varphi ({\bm v})$, as expected from equilibrium statistical mechanics. Secondly, it is seen that the density profile shows a coupling between the effect of both the external field and the strong confinement. When one moves along  the $z$ coordinate, perpendicularly to the hard walls, the density profile looks parabolic, as it happens in absence of external fields, although the amplitude depends on the  projected position on the walls, through the external potential. On the other hand, for a given $z$, the density profile parallel to the plates differs from that of a dilute $(d-1)$-dimensional gas in presence of the external field.

\section{Conclusions and final remarks}
\label{s5}
The primary objetive of this work has been to derive the kinetic theory description for a dilute gas of hard spheres or disks under strong confinement, starting  from the underlying Liouville equation, by means of a systematic analysis. The simple confining  geometry considered, two parallel plates separated a distance smaller than two particle diameters, permits to identify easily the right parameter  leading to a density expansion of the BBGKY hierarchy, obtained by direct integration of the Liouville equation. It is the ratio of the particles diameter to the mean free path associated to the projection of the trajectories of the particles on the plane or line parallel to the hard boundaries. Keeping only the two lowest orders in this parameter, a set of two closed partial differential equations is obtained for the one-particle distribution function. The method also provides all the reduced distribution functions, for arbitrary number of particles.  They define the Boltzmann level kinetic theory for the strongly confined system. This result differs significantly from both, the Boltzmann equation for bulk systems \cite{RydL77} and the recently proposed modified Boltzmann equation for strongly confined systems \cite{BMyG16,MBGyM22}. To put the relevance of the theory developed here in a proper context, the following points must be taken into account:
\begin{itemize}
\item The present kinetic theory  admits a stationary solution that is consistent with equilibrium statistical mechanics. In absence of external fields, the density profile is obtained only up to second order in the density and agree with the low density limit of the expression obtained by means of equilibrium density functional theory \cite{SyL96}.
\item When a stationary external potential perpendicular to the confinement direction is considered, the present kinetic theory also has a stationary solution consistent with equilibrium statistical mechanics, contrary to the modified Boltzmann equation. The density profile is more complicated than for a bulk dilute gas, since inhomogeneities coming from the confinement couple with those due to the external field, although if considered separately they induce inhomogeneities along perpendicular directions. This coupling is similar to what happens in a bulk system at high densities and it is a consequence of the collisions between particles.

\item The DSMC method have been used  to investigate the effect of the confinement on the transport properties of a gas \cite{LRyM16,ZMyW12}. Although the DSMC method follows the motion of the particles in the system, its idea is to try to mimic the dynamics of the particles as described by the assumed kinetic equation, i.e. it is based on the Boltzmann equation and, therefore, it can not provide any information about the validity of the kinetic equation itself. Nevertheless, it seems possible to adapt the usual DSMC method to the kinetic theory formulated here. 

\item The structure of the kinetic equation (\ref{3.14})  may induce to think that it is hard to apply to general non-equilibrium situations. Notice that no direct relationship between $f_{1}^{(0)}$ and $f_{1}^{(1)} $ is required by the theory, other than the kinetic equation itself. Nevertheless, on the basis of the symmetry of the several terms appearing in Eq. (\ref{3.14}), it can be expected that the equation actually decomposes into two equations in many situations, where some imposed symmetry is present. This is what happens in the equilibrium situations investigated in this paper.

\item As discussed in the Introduction section, one of the main reasons to develop the present kinetic theory is to investigate transport properties of the gas in the plane or line  parallel to the confining boundaries. For this purpose, it is convenient to consider an average distribution function defined as 
\begin{equation}
\label{5.1}
\overline{f}_{1}({\bm r}_{\perp},t)  \equiv \frac{1}{h-\sigma} \int_{\sigma /2}^{h-\sigma/2} dz\, f_{1}({\bm r},t).
\end{equation}
For this function, Eq. (\ref{3.14}) becomes (again, the characteristic functions and the tildes are omitted for simplicity)
\begin{eqnarray}
\label{5.2}
& \left( \frac{\partial}{\partial t} + {\bm v}_{1 \perp} \cdot {\bm \nabla}_{\perp} \right) \overline{f}_{1}^{(0)}({\bm r}_{\perp},t)+ \alpha v_{1z} \left[ f_{1}^{(1)} ({\bm r}_{\perp},h-\sigma/2,{\bm v}_{1},t) -    f_{1}^{(1)} ({\bm r}_{\perp},\sigma/2,{\bm v}_{1},t) \right] &   \nonumber \\
& = \frac{1}{h-z} \int_{\sigma /2}^{h-\sigma /2} dz_{1}\, J[x_{1} | \overline{f}_{1}^{(0)} ] &
 \end{eqnarray}     
 Notice that $\overline{f}_{1}^{(0)} = f_{1}^{(0)} $.
 Thus  only the boundary values of the original, non-averaged, distribution function are needed. Otherwise, one gets a closed equation for       $\overline{f}_{1}^{(0)}({\bm r}_{\perp},t)$.  Probably, in any practical applications the boundary term can be    safely neglected or modeled in an accurate form. For instance, it vanishes if a symmetric initial state with respect $z=0$ is considered. In any case, the above equation seems a good
 starting point to investigate transport and, in particular, hydrodynamics in strongly confined systems.    
 
 \item  Equation (\ref{5.2}) has been applied \cite{BGyM17,MBGyM22} to symmetric in $z$ systems, then becoming a close equation for $\overline{f}_{1}^{(0)}$, with an anisotropic initial velocity distribution. The relaxation of the velocity moments associated to the vertical and horizontal motions were analyzed in detail. As expected, they relaxed towards their common equilibrium value, and the theoretical result for the relaxation process was found to be in a very good agreement with the numerical data obtained by means of molecular dynamics simulations.   In these papers, the starting kinetic equation was obtained as an approximation, using heuristic arguments, while  Eq. (\ref{5.2}) has been derived here as an exact equation in the low density limit.

\item The extension of this kinetic theory to granular gases, i.e. systems of hard spheres or disks colliding inelastically, is straightforward. The only change to be done  affects to the collision rule defined in Eq. (\ref{2.1}) and, if needed, to the definition of the boundary conditions, e. g. energy can be injected into the system through them. It is clear that inelasticity does not affect the scaling carried out in Sec.\ \ref{s3} and, therefore, Eqs. (\ref{3.3}) and (\ref{3.4}), or (\ref{3.14}), remain valid in the low density limit for inelastic collisions. Quite interestingly, equations similar to (\ref{5.2}) have been employed in  the context of granular gases \cite{MGyB19,MGyB22}. While in those works the equation was considered as an approximation to Eq. (\ref{c1}), it has been shown in this paper that it is correct in the low density limit.

\end{itemize}

In summary, the analysis presented and the derived kinetic theory offers a firm basis for the description of transport in strongly confined gases of hard spheres or disks. This includes all the different regimes, from free flow to hydrodynamics, as well as inelastic or granular gases. Let us also mention that the extension to larger width  systems, of the order of a few particle diameters,  is straightforward \cite{MPGyM23}.

\section{Acknowledgements}

This research was supported by grant ProyExcel-00505 funded by Junta de Andaluc\'{\i}a and GrantPID2021-126348N funded by MCIN/AEI/10.13039/501100011033 and "ERDF A way of making Europe".

\appendix

\section{Density expansion of the BBGKY hierarchy}
\label{ap1}
As indicated in Sec. \ref{s3}, scaled positions and velocities are defined as
\begin{equation}
\label{a1}
x^{*}_{i} \equiv \left\{ {\bm r}^{*}_{i}, {\bm v}^{*}_{i} \right\}, 
\end{equation}
\begin{equation}
\label{a2}
{\bm r}^ {*}_{i}  = \frac{{\bm r}_{i \perp}}{\ell_{0}} + \frac{z_{i} \widehat{\bm e}_{z}}{\sigma} ,  \quad {\bm r}_{i \perp}= {\bm r}_{i}- z_{i} \widehat{\bm e}_{z}, \quad {\bm v}^{*}_{i}= \frac{{\bm v}_{i}}{v_{0}}\, ,
\end{equation}
while the dimensionless time scale is 
\begin{equation}
\label{a3}
t^{*} \equiv \frac{t v_{0}}{\ell_{0}}.
\end{equation}
Finally, the dimensionless  distribution functions are scaled by
\begin{equation}
\label{a4}
f^{*}_{s} (x_{1}^{*}, \ldots, x^{*}_{s},t^{*})= \left( \frac{v_{0}^{d}}{\tilde{n} \sigma^{-1}} \right)^{s} f_{s}(x_{1},\ldots,x_{s},t).
\end{equation}
In terms of the above dimensionless quantities, the BBGKY hierarchy, Eq. (\ref{2.27}), reads
\begin{eqnarray}
\label{a5}
\frac{\partial}{\partial t^{*}}f_{s}^{*} (x^{*}_{1}, \ldots, x^{*}_{s},t^{*} ) &+&\sum_{i=1}^{s}  \left( {\bm v}^{*}_{i \perp} \cdot {\bm \nabla}^{*}_{i\perp} \right) f_{s}^{*} (x^{*}_{1}, \ldots, x^{*}_{s},t^{*} )
+ \alpha^{-1} \sum_{i=1}^{s} v^{*}_{iz} \frac{\partial}{\partial z^{*}_{i}} f_{s}^{*} (x^{*}_{1}, \ldots, x^{*}_{s},t^{*} ) \nonumber \\
& -& \alpha^{-1} \sum_{i=1}^{s} \left[ \overline{T}^{(w)*}_{0}(i)+
\overline{T}^{(w)*}_{h}(i) \right] f_{s}^{*} (x^{*}_{1}, \ldots, x^{*}_{s},t^{*} ) \nonumber \\
&- & \alpha^{d-2} \sum_{1 \leq i < j \leq s} \overline{T}^{*} (i,j) f_{s}^{*} (x^{*}_{1}, \ldots, x^{*}_{s},t^{*} ) \nonumber \\
& = & \sum_{i=1}^{s} \int dx^{*}_{s+1}\, \overline{T}^{*}(i,s+1) f_{s+1}^{*} (x^{*}_{1}, \ldots, x^{*}_{s+1},t^{*} ),
\end{eqnarray}
where the density parameter
\begin{equation}
\label{a6}
\alpha \equiv \frac{\sigma}{\ell_{0}}= \widetilde{n} \sigma^{d-1}
\end{equation}
has been introduced and ${\bm \nabla}_{\perp}$ is used for the component of the gradient operator in the plane parallel to the hard boundaries. The dimensionless collision operators read
\begin{equation}
\label{a7}
\overline{T}_{0}^{(w)*} (i) = v_{iz}^{*} \delta \left( z^{*}_{i}-\frac{1}{2} \right) \left[ \theta (v^{*}_{iz}) b_{i}^{(w)*}-\theta (-v^{*}_{iz}) \right],
\end{equation}
\begin{equation}
\label{a8}
\overline{T}_{h}^{(w)*}(i) = -v^{*}_{iz} \delta \left( z^{*}_{i} - \frac{h}{\sigma} +\frac{1}{2} \right) \left[ \theta (-v^{*}_{iz}) b_{i}^{(w)*}+\theta (v^{*}_{iz}) \right],
\end{equation}
\begin{eqnarray}
\label{a9}
\overline{T}^{*}(i,j) & = &  \int d \widehat{\bm \sigma}\,  \theta ({\bm g}^{*}_{ij} \cdot \widehat{\bm \sigma}) | {\bm g}^{*}_{ij} \cdot \widehat{\bm \sigma} | 
\nonumber \\
&& \times \left[  \delta({\bm r}^{*}_{ij,\perp}- \alpha \widehat{\bm \sigma}_{\perp}) \delta (z^{*}_{ij} -\widehat{\sigma}_{z}  )b_{\bm \sigma}^{*}(i,j)-
\delta ({\bm r}^{*}_{ij,\perp}+ \alpha \widehat{\bm \sigma}_{\perp})  \delta ( z^{*}_{ij}+ \widehat{\sigma}_{z}) \right].
\end{eqnarray}
The operators $b_{i}^{(w)*}$ and $b_{\bm \sigma}^{*}(i,j)$ are defined as the  same operators without asterisk, Eqs. (\ref{2.1}) and (\ref{2.2}), but now acting on the scaled velocities.  Let us search for solutions of the above hierarchy given by a series expansion in powers of $\alpha$,
\begin{equation}
\label{a.10}
f_{s}^{*} (x^{*}_{1}, \ldots, x^{*}_{s},t^{*} ) = \sum_{k=0}^{\infty} \alpha^{k} f_{s}^{*(k)}(x_{1}, \ldots, x^{*}_{s},t^{*} ).
\end{equation}
When this is substituted into Eq. (\ref{a5}), to order $\alpha^{-1}$ it is found
\begin{equation}
\label{a11}
 \sum_{i=1}^{s} v^{*}_{iz} \frac{\partial}{\partial z^{*}_{i}} f_{s}^{*(0)} (x^{*}_{1}, \ldots, x^{*}_{s},t^{*} )  - \sum_{i=1}^{s} \left[ \overline{T}^{(w)*}_{0}(i)+
\overline{T}^{(w)*}_{h}(i) \right] f_{s}^{*(0)} (x^{*}_{1}, \ldots, x^{*}_{s},t^{*} ) =0.
\end{equation}
This hierarchy has the solution
\begin{equation}
\label{a12}
f_{s}^{*(0)} (x^{*}_{1}, \ldots, x^{*}_{s},t^{*} ) =\prod_{i=1}^{s}f_{1}^{*(0)}(x^{*}_{i},t^{*}),
\end{equation}
where $f_{1}^{*(0)}$ is a solution of
\begin{equation}
\label{a13}
v^{*}_{z}\frac{\partial}{\partial z^{*}}\, f^{*(0)}_{1}(x^{*},t^{*}) - \left[ \overline{T}^{(w)*}_{0}+
\overline{T}^{(w)*}_{h} \right] f_{1}^{*(0)} (x^{*},t^{*})=0.
\end{equation}
Because of Eq. (\ref{2.28}), one can write
\begin{equation}
\label{a14}
f^{*(0)}_{1} (x^{*},t^{*}) = W^{(w)*} (z^{*}) \widetilde{f}_{1}^{*(0)}(x^{*},t^{*}).
\end{equation}
The function $\widetilde{f}_{1}^{*(0)}(x^{*},t^{*})$ can be assumed to be continuous at the boundaries. Upon introducing this expression into Eq. (\ref{a13}), the equation decomposes into three equations, one for the regular part, another  for the singular part  at $z^{*}=1/2$, and
the last one for the singular part at $z^{*}= h/\sigma- 1/2$.  The three equations read
\begin{equation}
\label{a15}
W^{(w)*} (z^{*}) v_{z}^{*} \frac{\partial}{\partial z^{*}}\, \widetilde{f}_{1}^{*(0)}(x^{*},t^{*}) =0,
\end{equation}
\begin{equation}
\label{a16}
\theta(v^{*}_{z}) \delta \left(z^{*}- \frac{1}{2} \right) \widetilde{f}_{1}^{*(0)}(x^{*},t^{*}) =\theta(v^{*}_{z}) \delta \left(z^{*}- \frac{1}{2} \right) b^{(w)*} \widetilde{f}_{1}^{*(0)}(x^{*},t^{*}), 
\end{equation}
 \begin{equation}
 \label{a17}
 \theta(-v^{*}_{z}) \delta \left(z^{*}- \frac{h}{\sigma} + \frac{1}{2} \right) \widetilde{f}_{1}^{*(0)}(x^{*},t^{*}) =\theta(- v^{*}_{z}) \delta \left(z^{*}- \frac{h}{\sigma}+\frac{1}{2} \right) b^{(w)*} \widetilde{f}_{1}^{*(0)}(x^{*},t^{*}).
 \end{equation}
To derive Eqs.  (\ref{a16}) and (\ref{a17}), the explicit form of  the walls collisions operators, Eqs. (\ref{a7}) and (\ref{a8}), have been employed. These equations can be considered as boundary conditions. to Eq. (\ref{a15}). Actually, because of the latter, the boundary conditions can be expressed in the equivalent form
\begin{equation}
\label{a18}
\widetilde{f}_{1}^{*(0)}(x^{*},t^{*}) = b^{(w)*} \widetilde{f}_{1}^{*(0)}(x^{*},t^{*}),
\end{equation}
i.e., $\widetilde{f}_{1}^{*(0)}(x^{*},t^{*})$ is an even function of $v_{z}^{*}$.  Then, Eq.\, (\ref{a12}) implies that  all the zeroth-order reduced distribution functions $f_{s}^{*(0)} (x^{*}_{1},\ldots,x_{s}^{*})$ are even functions of all the $z$ components of the velocities $v^{*}_{iz}, i=1,\ldots,s$.

To zeroth order in $\alpha$, Eq. (\ref{a5}) leads to
\begin{eqnarray}
\label{a19}
\frac{\partial}{\partial t^{*}}\, f_{s}^{* (0)} (x^{*}_{1}, \ldots, x^{*}_{s};t^{*})  &+& \sum_{i=1}^{s} {\bm v}_{i \perp}^{*} \cdot \ {\bm \nabla}^{*}_{\perp} f_{s}^{* (0)} (x^{*}_{1}, \ldots, x^{*}_{s};t^{*})  
+ \sum_{i=1}^{s} v^{*}_{iz} \frac{\partial}{\partial z^{*}_{i}} f_{s}^{* (1)} (x^{*}_{1}, \ldots, x^{*}_{s};t^{*}) \nonumber \\ 
 & - &    \sum_{i=1}^{s} \left[ \overline{T}^{(w)*}_{0}(i)+
\overline{T}^{(w)*}_{h}(i) \right] f_{s}^{*(1)} (x^{*}_{1}, \ldots, x^{*}_{s},t^{*} )    \nonumber \\
& - & \delta^{kr}_{d,2}  \sum_{1 \leq i<j \leq s} \overline{T}_{0}^{*}(i,j) 
f_{s}^{*(0)} (x^{*}_{1}, \ldots, x^{*}_{s};t^{*})  \nonumber \\
& = & \sum_{i=1}^{s} \int dx_{s+1}  \overline{T}_{0}^{*} (i, s+1) f_{s+1}^{* (0)} (x^{*}_{1}, \ldots, x^{*}_{s+1};t^{*}),
\end{eqnarray}
where $\delta^{kr}_{n,m}$ is the Kronecker delta function and
\begin{equation}
\label{a19b}
\overline{T}_{0}^{*}(i,j)  =  \delta({\bm r}^{*}_{ij,\perp})  \int d \widehat{\bm \sigma}\,  \theta ({\bm g}^{*}_{ij} \cdot \widehat{\bm \sigma}) | {\bm g}^{*}_{ij} \cdot \widehat{\bm \sigma} |  \left[   \delta (z^{*}_{ij} -\widehat{\sigma}_{z}  )b_{\bm \sigma}^{*}(i,j)- \delta ( z^{*}_{ij}+ \widehat{\sigma}_{z}) \right].
\end{equation}
Consider first the case of a quasi-two-dimensional system of hard spheres ($d=3$). Taking into account Eq.\ (\ref{a12}), it is seen that the hierarchy has the solution
\begin{equation}
\label{a20}
f_{s}^{*(1)} (x^{*}_{1}, \ldots, x^{*}_{s},t^{*} ) = \sum_{i=1}^{s} \prod_{j \neq i}^{s} f_{1}^{*(1)} (x_{i}^{*},t^{*}) f_{1}^{*(0)} (x_{j}^{*},t^{*}),
\end{equation}
where $f_{1}^{*(1)}$ obeys a differential equation involving the walls collisions operators.  Similarly to Eq.\ (\ref{a14}), we can introduce $\widetilde{f}_{1}^{*(1)}$ by
\begin{equation}
\label{a21}
f^{*(1)}_{1} (x^{*},t^{*}) = W^{(w)*} (z^{*}) \widetilde{f}_{1}^{*(1)}(x^{*},t^{*}),
\end{equation}
so that one gets the equation
\begin{eqnarray}
\label{a22}
\lefteqn{ W^{(w)*} (z^{*}_{1})\left \{ \left( \frac{\partial}{\partial t^{*}} + {\bm v}_{1\perp}^{*} \cdot {\bm \nabla}^{*}_{1 \perp} \right) \widetilde{f}_{1}^{*(0)}(x^{*}_{1},t^{*})  + v_{1z}^{*} \frac{\partial}{\partial z_{1}^{*}} \widetilde{f}_{1}^{*(1)}(x_{1}^{*},t^{*}) \right\}  }\nonumber \\
&&= W^{(w)*} (z^{*}_{1}) \int dx^{*}_{2}\,  \overline{T}_{0}^{*} (1,2) \widetilde{f}_{1}^{*(0)}(x^{*}_{1},t^{*})  W^{(w)*} (z^{*}_{2}) \widetilde{f}_{1}^{*(0)}(x^{*}_{2},t^{*}) , 
\end{eqnarray}
to be solved with the boundary conditions
\begin{equation}
\label{a23}
\theta(v^{*}_{z}) \delta \left(z^{*}- \frac{1}{2} \right) \widetilde{f}_{1}^{*(1)}(x^{*},t^{*}) =\theta(v^{*}_{z}) \delta \left(z^{*}- \frac{1}{2} \right) b^{(w)*} \widetilde{f}_{1}^{*(1)}(x^{*},t^{*}), 
\end{equation}
 \begin{equation}
 \label{a24}
 \theta(-v^{*}_{z}) \delta \left(z^{*}- \frac{h}{\sigma} + \frac{1}{2} \right) \widetilde{f}_{1}^{*(1)}(x^{*},t^{*}) =\theta(- v^{*}_{z}) \delta \left(z^{*}- \frac{h}{\sigma}+\frac{1}{2} \right) b^{(w)*} \widetilde{f}_{1}^{*(1)}(x^{*},t^{*}).
 \end{equation}
  For a system of strongly confined hard disks ($d=2$), the one-particle distribution functions $\widetilde{f}_{1}^{*(0)}$ and $\widetilde{f}_{1}^{*(1)}$, defined by Eqs. (\ref{a14}) and (\ref{a21}) also satisfy Eqs. (\ref{a15}) and (\ref{a22}), respectively,  with the boundary conditions formulated in Eq.\ (\ref{a18}), and Eqs. (\ref{a23}) and (\ref{a24}). Also $f_{s}^{*(0)}(x^{*}_{1}, \ldots,x_{s}^{*},t^{*})$ is given by Eq. (\ref{a12}) for all $s$, but
 \begin{equation}
 \label{a25}
 f_{2}^{*(1)} (x^{*}_{1},x^{*}_{2},t^{*}) = f_{1}^{*(0)} (x_{1}^{*},t^{*}) f_{1}^{*(1)} (x_{2}^{*},t^{*} ) + f_{1}^{*(0)} (x_{2}^{*},t^{*}) f_{1}^{*(1)} (x_{1}^{*},t^{*} ) + g_{2}^{*(1)} (x^{*}_{1},x^{*}_{2},t^{*})
 \end{equation}
and
\begin{equation}
\label{a26}
f_{s}^{*(1)}(x^{*}_{1},\ldots,x^{*}_{s},t^{*}) = \sum_{i=1}^{s} \prod_{j \neq i}^{s}   f_{1}^{*(1)} (x_{i}^{*},t^{*}) f_{1}^{*(0)} (x_{j}^{*},t^{*} ) + \sum_{1 \leq i < j \leq s} \prod_{k \neq i,j}^{s} f_{1}^{*(0)} (x^{*}_{k},t^{*}) g^{*(1)}_{2}(x^{*}_{i},x^{*}_{j},t^{*})
\end{equation}
for $s \geq 3$. An equation for the correlation function $g^{*(1)}_{2}$ follows directly from the equations for $f_{2}^{*(1)}$, $f_{1}^{*(0)}$, and $f_{1}^{*(1)}$. Again, it  is convenient to introduce a function $\widetilde{g}_{2}^{*(1)}$ defined as
\begin{equation}
\label{a27}
 g^{*(1)}_{2}(x^{*}_{1},x^{*}_{2},t^{*}) = W^{(w)*}(z^{*}_{1}) W^{(w)*} (z^{*}_{2}) \widetilde{g}^{*(1)}_{2}(x^{*}_{1},x^{*}_{2},t^{*}).
 \end{equation}
Then, it is easily found that this function verifies the equation
\begin{eqnarray}
\label{a28}
 \lefteqn{W^{(w)}(z_{1}) W^{(w)}(z_{2}) \left( v_{1z}^{*} \frac{\partial}{\partial z_{1}^{*}} + v_{2z}^{*}  \frac{\partial}{\partial z_{2}^{*}} \right) \widetilde{g}^{*(1)}_{2}(x^{*}_{1},x^{*}_{2},t^{*})}  \nonumber \\
 & &= W^{(w)}(z_{1}) W^{(w)}(z_{2}) \overline{T}_{0}^{*} (1,2) \widetilde{f}_{1}^{*(0)} (x_{1}^{*},t^{*}) \widetilde{f}_{1}^{*(0)} (x_{2}^{*},t^{*}),
\end{eqnarray}
to be solved with the boundary conditions
\begin{eqnarray}
\label{a29}
 \lefteqn{\theta(v^{*}_{1z}) \delta \left( z^{*}_{1}- \frac{1}{2} \right)    W^{(w)*} (z^{*}_{2})\widetilde{g}_{2}^{*(1)} (x^{*}_{1},x^{*}_{2},t^{*}) } \nonumber \\
&&= \theta(v^{*}_{1z}) \delta \left( z^{*}_{1}- \frac{1}{2} \right) W^{(w)*} (z^{*}_{2})  b_{1}^{(w)*} \widetilde{g}_{2}^{*(1)} (x^{*}_{1},x^{*}_{2},t^{*}) ,
\end{eqnarray}
\begin{eqnarray}
\label{a30}
 \lefteqn{\theta(-v^{*}_{1z}) \delta \left( z^{*}_{1}- \frac{h}{\sigma} +\frac{1}{2} \right)    W^{(w)*} (z^{*}_{2})\widetilde{g}_{2}^{*(1)} (x^{*}_{1},x^{*}_{2},t^{*}) } \nonumber \\
&&= \theta(-v^{*}_{1z}) \delta \left( z^{*}_{1}- \frac{h}{\sigma} +\frac{1}{2} \right) W^{(w)*} (z^{*}_{2})  b_{1}^{(w)*} \widetilde{g}_{2}^{*(1)} (x^{*}_{1},x^{*}_{2},t^{*}),
\end{eqnarray}
and other two conditions obtained from the above by interchanging indices $1$ and $2$.

These results give an exact solution to the entire hierarchy to the two lowest orders in $\alpha$, with reduced distribution functions $f_{s}$ of any number of particles determined by the sum of products of single particles functions $f_{1}^{*(0)}(x^{*},t^{*})$  and $f_{1}^{*(1)}(x^{*},t^{*})$  and, for hard disks, of the pair correlation function $g_{2}^{*(1)}(x^{*}_{1},x^{*}_{2},t^{*})$. Equations (\ref{a22})-(\ref{a24}), when rewritten in terms of the original variables and functions, become Eqs. (\ref{3.4}), (\ref{3.6}), and (\ref{3.7}), while Eqs. (\ref{3.10})-(\ref{3.12}) can be recognized as Eqs. (\ref{a28})-(\ref{a30}).

\section{Another form of the collision term  for strongly confined dilute gases}
\label{ap2}
In this Appendix, the derivation of the modified Bolztmann theory for a strongly confined gas as given by  Eq. (\ref{3.14}) will be sketched. The starting point will be Eq. (\ref{3.4}). Consider the right hand side of this equation,
\begin{equation}
\label{b1}
I \equiv  W^{(w)} (z_{1}) \int dx_{2}\,  \overline{T}_{0} (1,2) \widetilde{f}_{1}^{(0)}(x_{1},t)  W^{(w)} (z _{2}) \widetilde{f}_{1}^{(0)}(x_{2},t).
\end{equation}
Using the expression of the operator $\overline{T}_{0}(1,2)$, Eq. (\ref{3.5}), carrying out the integral over ${\bm r}_{2 \perp}$, and changing from variable $\widehat{\bm \sigma}$ to $- \widehat{\bm \sigma}$ in the second adding inside the integral, one gets
\begin{eqnarray}
\label{b2}
I & = &\sigma^{d-1} W^{(w)}(z_{1}) \int dz_{2} \int d{\bm v}_{2} \int d \widehat{\bm \sigma}\, |{\bm g}_{12} \cdot \widehat{\bm \sigma} | \delta (z_{12} - \sigma_{z})
\left[ \theta ({\bm g}_{12}  \cdot \widehat{\bm \sigma}) b_{\bm \sigma} - \theta (-{\bm g}_{12}  \cdot \widehat{\bm \sigma}) \right] \nonumber \\
&& \times \widetilde{f}_{1}^{(0)}(x_{1},t)  W^{(w)} (z _{2}) \widetilde{f}_{1}^{(0)}(x_{2},t).
\end{eqnarray}
{\bf  a) Hard spheres} \\
Using spherical coordinates, it is $ \widehat{\bm \sigma} = \left( \sin \theta \sin \phi, \sin \theta \cos \phi, \cos \theta \right)$, $d \widehat{\bm \sigma} = \sin \theta d \theta d\phi $, and $\sigma_{z}= \sigma \cos \theta$, where $\theta$  and $\phi$ are the polar and azimuthal angles, respectively.  Then, a direct calculation gives
\begin{equation}
\label{b3}
\int d \widehat{\bm \sigma}\,  \delta (z_{12}-\sigma_{z}) F (\widehat{\bm \sigma},{\bm r}_{1}, {\bm r}_{2}) =
 \frac{1}{\sigma} \int_{0}^{2\pi} d \phi\, 
\left[ F (\widehat{\bm \sigma},{\bm r}_{1}, {\bm r}_{2} ) \right]_{\cos \theta = z_{12} / \sigma} \, ,
\end{equation}
for arbitrary $F$. Using this into Eq. (\ref{b2}) and applying the definition of the characteristic function $W^{(w)}(z_{2})$, one gets
\begin{equation}
\label{b4}
I= W^{(w)}(z_{1} ) J[x_{1}|f_{1}^{(0)}],
\end{equation}
where $J[x_{1}|f_{1}]$ is defined by Eq. (\ref{3.15}). Substitution of the above expression for $I$ into Eq.\ (\ref{3.4}) directly gives the modified Boltzmann theory, Eq. (\ref{3.14}) for strongly confined hard spheres. The positive sign for $\sin \theta$  in Eq.\ (\ref{3.17}) is a consequence of the interval of variation of the polar angle, $0 \leq\ \theta  \leq \pi $.

\noindent {\bf  b) Hard disks} \\
For $d=2$, in polar coordinates it is $\widehat{\bm \sigma} = ( \cos \phi, \sin \phi)$ and $d \widehat{\bm \sigma} = d\phi$, where $\phi$ is again the polar angle. Consider the expression $I$ defined in Eq. (\ref{b2}). The evaluation of the integral given in Eq. (\ref{b3}) requires now a decomposition of the integration region as
\begin{eqnarray}
\label{b5}
\lefteqn{\int d \widehat{\bm \sigma}\,  \delta (z_{12}-\sigma_{z}) F (\widehat{\bm \sigma},{\bm r}_{1}, {\bm r}_{2})} \nonumber \\
&&= \int_{-\pi/2}^{\pi /2} d\phi\, \delta ( z_{12}-\sigma \sin \phi)   F (\widehat{\bm \sigma},{\bm r}_{1}, {\bm r}_{2})    + \int_{\pi/2}^{3\pi/2} d\phi\,  \delta ( z_{12}-\sigma \sin \phi)   F (\widehat{\bm \sigma},{\bm r}_{1}, {\bm r}_{2}),
\end{eqnarray}
so that it is obtained
\begin{equation}
\label{b6}
I =  \frac{1}{\sigma \sqrt{1- \left( \frac{z_{12}}{\sigma} \right)^{2}}} F (\widehat{\bm \sigma}_{1},{\bm r}_{1}, {\bm r}_{2})  
+ \frac{1}{\sigma \sqrt{1- \left( \frac{z_{12}}{\sigma} \right)^{2}}} F (\widehat{\bm \sigma}_{2},{\bm r}_{1}, {\bm r}_{2}). 
\end{equation}
The unit vectors $\widehat{\bm \sigma}_{1}$ and $\widehat{\bm \sigma}_{2}$ are given by
\begin{equation}
\label{b7}
\widehat{\bm \sigma}_{1} = \left( \frac{z_{12}}{\sigma}, \sqrt{1- \left( \frac{z_{12}}{\sigma} \right)^{2}} \right), \quad 
\widehat{\bm \sigma}_{2} = \left( \frac{z_{12}}{\sigma}, - \sqrt{1- \left( \frac{z_{12}}{\sigma} \right)^{2}} \right)
\end{equation}
When this result is introduced into Eq. (\ref{b2}) and the function $W^{(w)}$ is used to delimite the integral over $z$ range, Eq. (\ref{3.18}) is obtained.

\section{Previous proposals of kinetic equations for quasi-one dimensional and quasi-two dimensional system of hard particles}
\label{ap3}

For  strongly confined systems of hard spheres or disks,  modified Boltzmann equations have been proposed \cite{BMyG16,BGyM17}. They have the form
\begin{equation}
\label{c1}
\left( \frac{\partial}{\partial t} + {\bm v}_{1} \cdot {\bm \nabla}_{1} \right) f_{1}({\bm r}_{1},{\bm v}_{1}, t) = J_{c}[{\bm r}_{1}, {\bm v}_{1}|f_{1}],
\end{equation}
where the collision term is
\begin{eqnarray}
\label{c2}
 J_{c}[{\bm r}_{1}, {\bm v}_{1}|f_{1}] &= & \sigma \int_{\sigma/2}^{h- \sigma/2} dz_{2} \int d{\bm v}_{2} \int_{0}^{2 \pi} d \phi\, | {\bm g}_{12} \cdot \widehat{\bm \sigma} |   \left[ \theta ( {\bm g}_{12} \cdot \widehat{\bm \sigma} )  b_{\bm \sigma} (12) - \theta ( -{\bm g}_{12} \cdot \widehat{\bm \sigma}) \right]  \nonumber \\
&& \times f_{1}({\bm r}_{1},{\bm v}_{1} ,t) f_{1}({\bm r}_{1\perp}, z_{2},{\bm v}_{2},t)
\end{eqnarray}
for hard spheres \cite{BMyG16,BGyM17}, while
\begin{eqnarray}
\label{c.3}
J_{c}[{\bm r}_{1}, {\bm v}_{1}|f_{1}] & = & \int_{\sigma/2}^{h- \sigma/2} dz_{2} \int d{\bm v}_{2} \frac{1}{\cos \theta_{1}}\, \left( 1 + \mathcal{P}_{{\bm \sigma}_{1}, {\bm \sigma}_{2}} \right) | {\bm g}_{12} \cdot \widehat{\bm \sigma}_{1} |   
\nonumber \\
&& \times\left[ \theta ( {\bm g}_{12} \cdot \widehat{\bm \sigma}_{1} )  b_{\bm \sigma} (12) - \theta ( -{\bm g}_{12} \cdot \widehat{\bm \sigma}_{1|}) \right]  
 f_{1}({\bm r}_{1},{\bm v}_{1} ,t) f_{1}(x_{1},z_{2},{\bm v}_{2},t).
 \end{eqnarray}
for a system of hard disks \cite{MBGyM22}. Of course, the kinetic equations hold for non-overlapping particles inside  the system, and must be solved with the appropriatted boundary conditions, namely those used along this paper. Therefore, the characteristic function $W^{(w)}(z_{1})$ appearing in Eq. (\ref{3.14}) is also implicit in Eq. (\ref{c1}).


\begin{thebibliography}{99}
\bibitem{note} A discussion of the Enskog equation for confined systems  is presented in \cite{MGyB18}.
\end{thebibliography}
 \end{document}